\documentclass[12pt,english]{article}
\usepackage{a4wide,latexsym}
\usepackage{babel}
\usepackage{epsfig}
\usepackage[latin1]{inputenc}

\newtheorem{theorem}{Theorem}
\newtheorem{lemma}{Lemma}
\newtheorem{algorithm}{Algorithm} 
\newenvironment{proof}{{\it Proof:\/}}{$\Box$\vskip 0.1in}

\begin{document}

  %log (keep uptodate at each major modification):\\
  %durr - jeu aoû 29 17:53:41 MEST 2002\\
  %durr - jeu aoû 29 18:14:48 MEST 2002\\
  %durr - Wed Sep 4 11:20:02 MEST 2002\\
  %durr - Tue Nov 12 16:22:05 MET 2002\\
  %durr - Tue Dec 10 09:59:56 PST 2002\\
  %durr - Wed Dec 11 18:44:18 PST 2002\\
  %durr - Tue Feb 18 16:47:22 MET 2003
  %mehdi - 25 fev 2003
  %mehdi -3,4 mars 2003
  %durr - Wed Mar  5 13:23:17 MET 2003
  %backtrace in the version tree - can't get rid of log n factor
  %durr - Fri Mar 21 16:05:10 CET 2003
  %mehdi - (add quant-ph reference)
  %durr - Tue Mar 25 12:12:18 CET 2003
  %durr - mar mar 25 23:56:42 CET 2003
  %durr - mer mar 26 22:54:20 CET 2003
  %durr - Thu Mar 27 08:47:51 MET 2003
  %durr - Thu Mar 27 16:49:49 CET 2003
  %durr - jeu mar 27 22:07:37 CET 2003
  %durr - dim mar 30 00:36:11 CET 2003
  %durr - jeu avr  3 14:42:45 CEST 2003

  \title{Quantum query complexity of graph connectivity}

  \author{
Christoph Dürr\thanks{%
Laboratoire de Recherche en Informatique, Université Paris-Sud,
91405 Orsay, France. durr@lri.fr. Research partially supported
by the EU 5-th framework programs QAIP IST-1999-11234, RESQ
IST-2001-37559, and RAND-APX IST-1999-14036, by CNRS/STIC 01N80/0502
and 01N80/0607 grants, by ACI Cryptologie CR/02 02 0040 grant of
the French Research Ministry.}
				\and
Mehdi Mhalla\thanks{%
Laboratoire Leibniz, Institut IMAG, Grenoble, France. 
Mehdi.Mhalla@imag.fr.}
				\and
Yaohui Lei\thanks{%
Département d'Informatique et recherche opérationnelle,
Université de Montréal,
CP 6128 succ Centre Ville,
Montréal QC     H3C 3J7,
Canada. leiyaohu@iro.umontreal.ca.}}

  \maketitle

  \begin{abstract}
    Harry Buhrman et al gave an $\Omega(\sqrt n)$ lower bound for
    monotone graph properties
    in the adjacency matrix query model.  
    Their proof
    is based on the polynomial method.  
    However for some properties stronger lower bounds exist.
    We give an $\Omega(n^{3/2})$
    bound for \textsc{Graph Connectivity}
    using Andris Ambainis' method, and an $O(n^{3/2}\log n)$
    upper bound based on Grover's search algorithm.  In addition we
    study the adjacency list query model, where we have almost
    matching lower and upper bounds for \textsc{Strong Connectivity} of
    directed graphs. 
  \end{abstract}

  \section{Introduction}

  The goal of the theory of quantum complexity is to find out what the
  computational speedup of a quantum computer is with respect to
  classical computers. Today there are only a few results which give a
  polynomial time quantum algorithm for some problem for which no
  classical polynomial time solution is known. We are interested in
  studying the speedup for problems for which there is already an
  efficient classical algorithm, but which could still be slightly
  improved quantumly.  Basic graphs problems are interesting
  candidates.  For example Heiligman studied the problem of computing
  distances in graphs~\cite{H03}.

  This paper addresses \textsc{Graph Connectivity}. We are given a graph
  $G(V,E)$ and would like to know whether for every vertex pair
  $(s,t)$ there is a path connecting them. A generalization is \textsc{Strong
  Connectivity} where we are given a \emph{directed} graph $G(V,E)$. To
  be strongly connected there must be a path from $s$ to $t$ but also
  from $t$ to $s$. Let's assume the vertex set
  $V=\{v_0,v_1,\ldots,v_{n-1}\}$.
  We call an undirected graph $k$-regular if the degree of all 
  its vertexes is $k$.  We say that a directed graph has outdegree $k$
  if all its vertexes have outdegree $k$.  If this also true for the indegree
  then we call it $k$-regular.

  Since it is extremely difficult to prove lower bounds for time
  complexity, we study the query complexity of these problems; meaning
  the minimal number of queries to the graph in order to solve the
  problem.  There are essentially two query models for graphs~:
  \begin{description}
  \item[The matrix model,] where the graph is given as the adjacency
    matrix $M\in\{0,1\}^{ n\times n} $, with $M_{ij}=1$ iff $(v_i,v_j)
    \in E$,
  \item[and the list model.]  In the later case we consider for this
    paper only directed graphs with fixed out-degree $k$.
    The encoding is a function $ f:[n]\times [k]\rightarrow [n]
    $ such that $ f(u,i) $ is the i-th neighbor of $ u $
    --- using an arbitrary numbering from $0$ to $k-1$ of the outgoing
    edges ---  and which
    satisfy the \emph{simple graph} promise
    \begin{equation}
      \label{eq:hyper}
      \forall u\in [n],i,j\in [k],i\neq j :
      f(u,i)\neq f(u,j)
    \end{equation}
  \end{description}
  which ensures that the graph is not a multigraph, i.e. does not have
  multiple edges between two nodes.  For undirected graphs we require
  an additional promise on the input, namely that $M$ is symmetric in the
  matrix model, and for the list model $\forall u,v\in [n]$ if
  $\exists i\in[k]: f(u,i)=v$ then $\exists j\in[k]:f(v,j)=u$.  Note
  that \textsc{Connectivity} is not what is called a 
  \emph{promise problem}, as the promise is input independent.

  The classical randomized query complexity in the matrix model is
  well studied. While it is believed to be $ \Omega (n^{2}) $, there
  is no matching lower bound at the moment. The first good lower bound
  of $ \Omega (n^{4/3}) $ is by Hajnal~\cite{Ha91}, and has recently
  be improved to $ \Omega (n^{4/3}(\log n)^{1/3}) $ by Chakrabart and
  Khot~\cite{CK01} and one year later even to $ \Omega (n^{2}/\log n)
  $ in a neat paper by Friedgut, Kahn and Wigderson~\cite{FKW}. We are
  not aware of any lower bound for \textsc{Strong Connectivity}.

  \begin{table}[htb]
    \begin{center}
      \begin{tabular}{|l|c|} \hline
	&(un)directed graphs\\ 
	\hline 
	matrix model & $\Omega(n^{3/2})$, $O(n^{3/2}\log  n)$\\ \hline
	matrix model, $k$-regular  & $O(n^{3/2}\sqrt k)$\\ \hline 
        \multicolumn{2}{c}{} \\
        \hline
	& directed graphs\\ \hline
	list model   & $\Omega(n\sqrt k)$, $O(n\sqrt k \log n)$\\ \hline
	list model, $k$-regular  &$\Omega(n)$\\
	\hline
      \end{tabular}
    \end{center}
    \caption{Query complexity of \textsc{(Strong) Connectivity}.}
  \end{table}

  The quantum query complexity of any monotone graph property
  in the matrix model was shown to be $
  \Omega (\sqrt{n}) $ by Buhrman, Cleve, de Wolf and Zalka~\cite{BCWZ}
  using the polynomial method and conjectured to be $\Omega(n)$. In
  this note we give the lower bound $\Omega(n^{3/2})$
  for \textsc{Connectivity}, a particular monotone graph 
  property.\footnote{In a previous version of this paper, we said that
  Buhrman et al conjectured $\Omega(n)$ for \textsc{Connectivity}, and
  since their conjecture concerns arbitrary monotone graph properties
  in general, we gave a false impression of improving their result. 
  We apologize.}
  We also have
  an almost matching upper bound of $O(n^{3/2}\log n)$ and for the
  special case when the out-degree is $k$ there is a trivial algorithm
  in $O(n^{3/2}\sqrt k)$ which is of interest when $k\in o(\log^2 n)$.
  These bounds do also hold for \textsc{Strong Connectivity}.

  For the list model we are not aware of any classical lower bound.
  Trivial classical upper bound is $O(nk)$. We show that the quantum
  query complexity 
  of \textsc{Strong Connectivity} 
  is $\Omega (n\sqrt{k})$ and $O(n\sqrt k \log n)$,
  which becomes better than classical for $k\in \Omega(\log^2 n)$.
  However our lower bound does not hold anymore for 
  the special case of $k$-regular
  directed graphs, for which we only have $\Omega(n)$.

  The time complexity of our algorithms are essentially the same as
  the query complexity.  However a $\log n$ factor applies in the bit
  model, when each vertex is encoded using $\log n$ bits.

  The space requirement is $O(\log n)$ qubits and $O(n\log n)$
  classical bits.  If we constraint the space (both classical and
  quantum) to $O(\log n)$ qubits then an algorithm is basically
  restricted to make an oblivious random walk on the graph.  Quantum
  random walks has been subject of several
  papers~\cite{AAKV01,CCDFGS02,K03}, in particular for the
  \textsc{$st$-Connectivity} problem~\cite{Wa01}.

  \section{The upper bounds for \textsc{Connectivity}}

  The upper bounds gain their speedup of classical algorithm only from
  Grover's search algorithm. Since it is an important tool for this
  paper we restate the exact results. The algorithm solves the search
  problem of finding an index $i$ which a given boolean table $T$ maps
  to $1$. Let $n$ be the size of $T$ and $t$ the number
  of $1$-entries in $T$.  In this paper we will use three versions of
  the search algorithm.
  \begin{itemize}
  \item 
    When the number of solutions $t$ is known in advance, Grover's
    search algorithm returns a solution after $O(\sqrt{n/t})$ queries
    to $T$ with 
    probability of error $O(t/n)$~\cite{Gro96}.
  \item
    When the number of solutions is not known in advance, then an
    extension of Grover's algorithm can be used, by Boyer et al which
    makes $O(\sqrt n)$ queries to $T$ and either outputs a solution or
    claims that $T$ is all $0$.  This second algorithms however errs
    with constant probability~\cite{BBHT}.
  \item
    There is another extension of the previous algorithm which finds
    the smallest index $i$ such that $T[i]=1$, in time $O(\sqrt n)$.
    Again error probability is constant~\cite{DH96}.
  \end{itemize}

  The main idea for the upper bound follows immediately from using
  Grover's search algorithm as a blackbox in an algorithm for
  constructing a spanning tree.

  \begin{theorem}
    The quantum query complexity of \textsc{Connectivity}
    is $O(n^{3/2}\log n)$
    in the adjacency matrix query model and $O(n \sqrt k \log n)$ in
    the adjacency list query model.
  \end{theorem}
  \begin{proof}
    We have to decide whether from vertex $v_0$, all other vertexes
    can be reached.  For that purpose we construct a depth first
    spanning tree of the connected component containing $v_0$.  We
    maintain a stack of vertexes which neighborhood still remains to
    be explored. Initially it contains only $v_0$. Whenever a new
    vertex is found, it is marked and pushed on the stack. At every
    time, we search for unmarked neighbors of the top vertex. If there
    is no, we remove the vertex from the stack.

    \begin{algorithm}[Depth-first search]
      \begin{tabbing}\\
	Let $A$ be an empty edge set, which will contain a spanning tree.\\
	Initially set of marked vertexes $ S=\{v_0\} $\\
	and stack of vertexes to be processed $ T=\{v_0\} $.\\
	While \=$ T\neq \{\} $ do\\
	\>	choose  topmost $ u\in T $ \\
	\>	use Grover's algorithm to find a neighbor $ v $ of $ u $ and not in $ S $\\
	\>	if success add $ v $ to $ S $, $(u,v)$ to $A$ and push $v$ on $ T $\\
	\>	otherwise pop $ u $ from $ T $.\\
	Answer "Yes the graph is connected" if all vertexes are marked ($S=V$).
      \end{tabbing}
      \label{algo:grover}
    \end{algorithm}

    To make the total success probability constant, we need each of the
    at most $2n$ Grover's searches to success with probability 
    $p$, such that $p^{2n}=2/3$.  This can be achieved using $O(\log n)$ 
    repetitions.\footnote{Ronald de Wolf informed us that only $\sqrt{\log n}$ 
    repetitions are necessary, since 
    in~\cite{BCWZ} it is shown that $\sqrt{n\log(1/\epsilon)}$ queries are 
    necessary to obtain an error probability $\epsilon$ in Grover's search algorithm.}  
    Therefore the cost of adding or removing a
    vertex from the stack is $O(\sqrt{n}\log n)$ in the matrix model and $
    O(\sqrt{k}\log n) $ in the list model.  This gives us the required
    upper bound.
  \end{proof}

  For graphs with small out-degree $k\in o(\log^2 n)$, there is
  a trivial but better algorithm for the matrix model.  
  Its complexity is $O(n^{3/2}\sqrt k)$.
  We use the fact,
  that Grover's search procedure behaves much better, when the number
  of solutions is known in advance.
  Let $m=nk$ be the number of edges in the graph.

  We claim that the following algorithm has the required complexity.

  \begin{algorithm}[Learning the adjacency matrix] \label{alg:learning}
    \begin{tabbing}\\
      Empty adjacency list $L$.\\
      Matrix $M'\in\{0,1\}^{n\times n}$ initially all zero.\\
      For \=$t=m$ downto $1$\\
      \>Quantum search $i,j$ s.t. $M_{ij}=1$ and $M'_{ij}=0$ 
      until a solution is found\\
      \>(using the fact that there are $t$ solutions in a 
           search space of size $n^2$)\\
      \>add $(i,j)$ to $L$ and set $M'_{ij}=1$\\
      Run classical spanning tree algorithm using adjacency list $L$.
    \end{tabbing}
    \label{algo:learning}
  \end{algorithm}

  The expected number of queries to $M$ is $O(\sqrt{n^2/m} +
  \sqrt{n^2/(m-1)} + \ldots + \sqrt {n^2})=O(n\sum_{x=1}^m x^{-1/2})$.
  We bound the sum by
\[
	\sum_{x=1}^m x^{-1/2} 
              \le\int_{x=1}^m x^{-1/2} 
                = \left[ 2x^{1/2} \right]_1^m \in O(\sqrt m). 
\]
  So in expected time $O(n
  \sqrt m)$ we know the complete graph, and can compute classically the
  solution. Stopping the overall algorithm at twice the expected time
  gives an algorithm with success probability at least $1/2$ and worst
  case running time $O(n \sqrt m)$.

  \section{The upper bound for \textsc{Strong Connectivity}}

  A directed graph is strongly connected if for a fixed vertex $v_0$,
  (1) all other vertexes can be reached by $v_0$, and (2) $v_0$ can be
  reached by all other vertexes.  The algorithms we gave in the
  previous section verifies condition (1), and in the matrix model,
  condition (2) can easily be verified by applying the same algorithm
  on the transposed adjacency matrix, i.e. taking all edges in reverse
  order.

  For the adjacency list model however more work is needed.

  \begin{theorem}
    The quantum query complexity of \textsc{Strong Connectivity} 
    for directed
    graphs with out-degree $k$ in the list model is $O(n \sqrt k \log
    n)$.
  \end{theorem}
  \begin{proof}
    In a first stage we use algorithm~\ref{algo:grover} to construct a
    directed spanning tree $A \subseteq E$ rooted in $v_0$.  Assume
    vertexes to be named according to the order
    algorithm~\ref{algo:grover} marked them, so $v_0$ was the first to
    be marked, $v_1$, the second, and so on.

    Then in a second stage we search for every vertex $v_i\in V$, the
    neighbor $v_j$ with smallest index.  The result is a set of
    backward edges $B\subseteq E$.  This search can be done with
    $O(n\sqrt k\log n)$ queries using Dürr and Høyer's variant of
    Grover's algorithm~\cite{DH96}.  We claim that the graph $G(V,E)$
    is strongly connected iff its subgraph $G'(V,A \cup B)$ is
    strongly connected, which would conclude the proof.

    Clearly if $G'$ is strongly connected then so is $G$ since $A\cup
    B\subseteq E$.  Therefore to show the converse assume $G$ strongly
    connected.  For a proof by contradiction let $v_i$ be the vertex with
    smallest index, which is not connected to $v_0$ in $G'$.  However
    by assumption there is a path in $G$ from $v_i$ to $v_0$.  Let
    $(v_l, v_{l'})$ be its first edge with $l\ge i$ and $l'<i$.  We
    will use the following property of depth first search.
    \begin{quote}
      \begin{lemma}
	\label{lem:prof}
	Let $v_l$ and $v_{l'}$ be two vertexes in the graph $G$ with
	$l<l'$. If there is a path from $v_l$ to $v_{l'}$ in $G(V,E)$
	then $v_{l'}$ is in the subtree of $G''(V,A)$ with root $v_l$.
      \end{lemma}
    \end{quote}

    Therefore we can replace in the original path the portion from
    $v_i$ to $v_l$ by a path only using edges from $A$.  Let $v_{l''}$
    be the neighbor of $v_l$ with smallest index.  Clearly $l''\leq l'
    <i$.  By the choice of $v_i$, there exists a path from $v_{l''}$
    to $v_0$ in $G'$.  Together this gives a path from $v_i$ to $v_0$
    in $G'$ contradicting the assumption and therefore concluding the
    proof.
  \end{proof}

  \section{The lower bounds}

  We will first give a simple lower bound for \textsc{Connectivity} (and 
  \textsc{Strong
  Connectivity}) in the list model, by a reduction from
  \textsc{Parity}.  As we recently found out, this reduction has first
  been used by Henzinger and Fredman for the on-line connectivity
  problem~\cite{HF98}.  We show later how to improve this
  construction.
  \begin{lemma}
    \textsc{Strong Connectivity} needs $ \Omega (n) $ queries in the list model.
    \label{lem:parity}
  \end{lemma}
  \begin{proof}
    We will use a straightforward reduction from \textsc{Parity}. 

    \begin{figure}[ht]
      \centerline{\epsfig{file=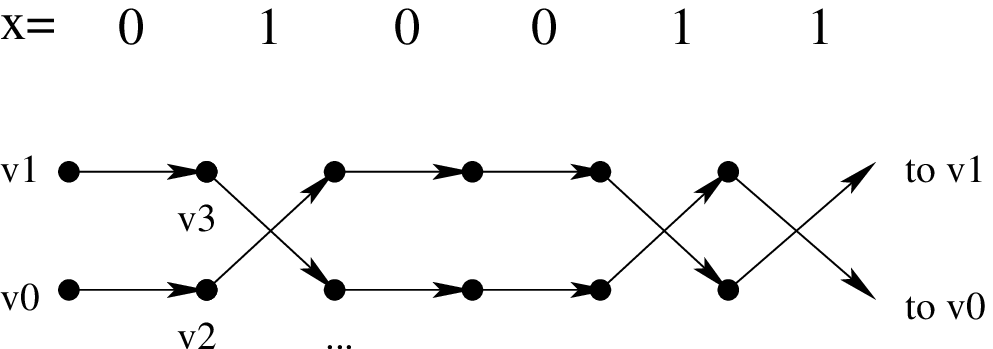,width=8cm}}
      \caption{A standard reduction from parity}
      \label{fig:parity}
    \end{figure}

    Let $ x\in \{0,1\}^{p} $ be an instance to the parity problem. We
    construction a permutation $ f $ on $ V=\{v_0,\ldots,v_{2p-1}\} $ which has exactly 1
    or 2 cycles depending the parity of $ x $. For any $ i\in [p] $
    and the bit $b=x_i$ we define $ f(v_{2i})=v_{2i+1+b}$ and 
    $ f(v_{2i+1})=v_{2i+3-b}$ where addition is modulo $2p$.
    See figure~\ref{fig:parity}. 
    The graph defined by $ f $ has 2 levels and $ p $ columns,
    each corresponding to a bit of $ x $. A directed walk starting at
    vertex $ v_0 $ will go from left to right, changing level whenever
    the corresponding bit in $ x $ is 1. So when $ x $ is even the
    walk returns to $ v_0 $ while having explored only half of the
    graph, otherwise it returns to $ v_1 $ connecting from there again
    to $ v_0 $ by $ p $ more steps.  Since the query complexity of
    \textsc{Parity} is $\Omega(n)$ --- see for example \cite{BCWZ} ---
    this concludes the proof.
  \end{proof}

  The following proofs will all rely on Andris Ambainis' technique for
  proving lower bounds, which we restate here.

  \begin{theorem}[{Ambainis~\cite[theorem 6]{AM02}}] 
    Let $L\subseteq \{0,1\}^*$ be a decision problem. Let $X\subseteq
    L$ be a set of positive instances and $Y\subseteq \bar L$ a set of
    negative instances. Let $R\subseteq X\times Y$ be a relation between
    instances of same size.  Let be the values
    $ m,m',l_{x,i} $ and $ l'_{y,i} $ with
    $x,y\in\{0,1\}^n$ and $x\in X$, $y\in Y$, $i\in[n]$ such that
    \begin{itemize}
    \item for every $x\in X$ there are at least $m$ different $y\in Y$
      in relation with $ x $, 
    \item for every $y\in Y$ there are at least $m'$ different $x\in X$
      in relation with $y$, 
    \item for every $x\in X$ and $i\in[n]$ there are
      at most $l_{x,i} $ different $y\in Y$ in relation with
      $x$ which differ from $x$ at entry $i$, 
    \item for every $y\in Y$ and $i\in [n]$ there are
      at most $l'_{y,i} $ different $x\in X$ in relation with
      $y$ which differ from $y$ at entry $i$. 
    \end{itemize}
    Then the quantum query complexity of $L$ is
    $\Omega (\sqrt{mm'/l_{\max }}) $
    where $ l_{\max }=\max _{x,y,i}l_{x,i}l'_{y,i}$
    with the maximum taken over $x R y$. 
  \end{theorem}

  Now we show how to improve our previous proof by changing slightly
  the construction.

  \begin{theorem}
    \textsc{Strong Connectivity} for directed graphs with out-degree $ k $
    needs $ \Omega (n\sqrt{k}) $ queries in the list
    model.\label{thm:hyper}
  \end{theorem}
  \begin{proof}
    We use a similar construction as for Lemma~\ref{lem:parity}, 
    but now for every vertex the
    $k-1$ additional edges are redirected back to an origin.
    We would like to connect them back to a fixed vertex
    $u_0$, but this would
    generate multiple edges and we want the proof work for simple graphs.
    Therefore we connect them back to a $k$-clique
    which then is connected to $u_0$.  See figure~\ref{fig:origin}. 
    Let be the vertex set 
    $V=\{v_{0},\ldots ,v_{2p-1},u_{0},\ldots ,u_{k-1}\} $ for some
	integer $p$.
    In the list model, the edges are defined by a function $f:V\times
    [k]\rightarrow V $.
    We will consider only functions with the following restrictions:

    For every $i\in [k]$ we have $f(u_i,0)=v_0$ and for 
    $j\in\{1,\ldots,k-1\}$ $f(u_i,j)=u_{i+j}$, 
    where addition is modulo $k$.

    For every $i\in [p]$ there exist $j_0, j_1\in [k]$ and a bit $b$ 
    such that $f(v_{2i},j_0)=v_{2i+2+b}$ 
    and $f(v_{2i+1},j_1)=v_{2i+3-b}$, where addition is 
    modulo $2p$ this time.  We call these edges the 
    \emph{forward edges}.  The \emph{backward edges}
    are for all $j\in[k]$ 
    $f(v_{2i},j)=u_j$ whenever $j\neq j_0$ and 
    $f(v_{2i+1},j)=u_j$ whenever $j\neq j_1$.

    \begin{figure}[ht]
      \centerline{\epsfig{file=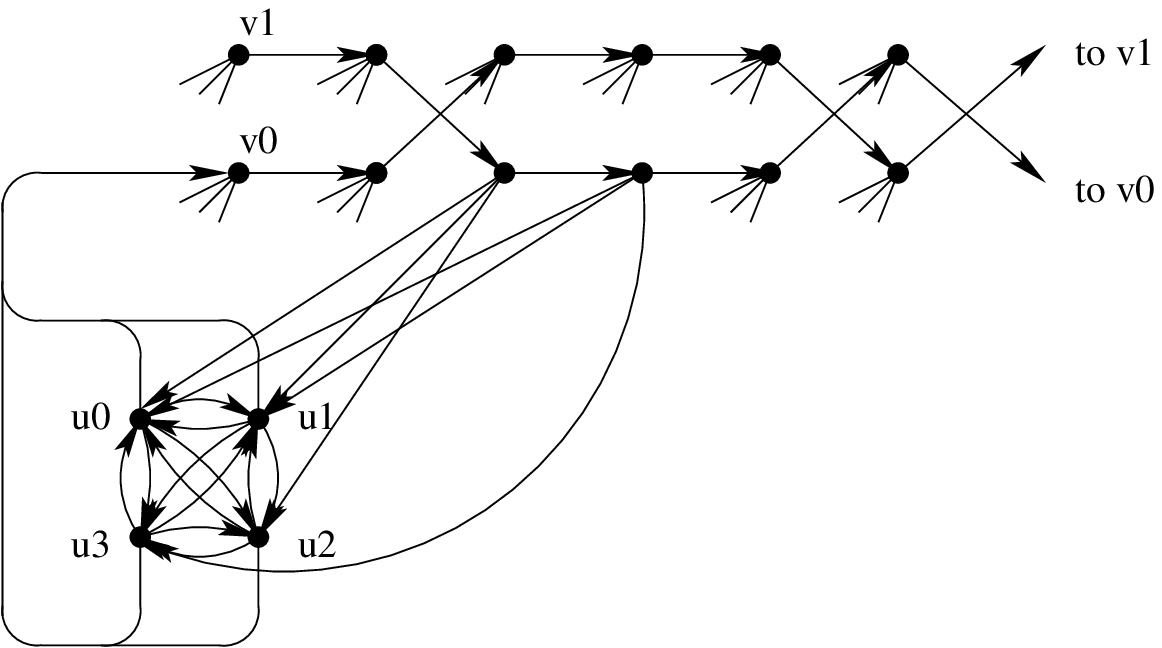,width=10cm}}
      \caption{A strongly connected graph}
      \label{fig:origin}
    \end{figure}

    Now all the nodes are connected to the $k$-clique, 
    the clique is connected to $ v_{0} $, 
    and the graph is strongly connected if and only if
    there is a path from $ v_{0} $ to $ v_{1} $.

    Let $X$ be the set of functions 
    which define a strongly connected graph,
    and $Y$ the set of functions which do not.
    Function $ f\in X $ is in relation with
    $g\in Y$ if there are 
    numbers $ i\in[p],j_{0},j_{1},h_{0},h_1\in[k]$ 
    with $j_0\neq h_0$, $j_1\neq h_1$
    such that the only
    places where $ f $ and $ g $ differ are
    \begin{eqnarray*}
      g(v_{2i},h_{0})=f(v_{2i+1},j_{1}) &  
                    & g(v_{2i+1},h_{1})=f(v_{2i},j_{0})		\\
      g(v_{2i},j_{0})=u_{j_0}&
                    & g(v_{2i+1},j_{1})=u_{j_1} 		\\
      f(v_{2i},h_{0})=u_{h_0} &  
                    & f(v_{2i+1},h_{1})=u_{h_1}
    \end{eqnarray*}
    Informally $f$ and $g$ are in relation if there is a level, where
    the forward edges are exchanged between a \emph{parallel} and 
    \emph{crossing} configuration and in addition the edge labels
    are changed.

    Then $m=m'=O(nk^{2})$, $p\in O(n)$ for the number of
    levels and $(k-1)^2$ for the number of possible
    forward edge labels. 
    We also have $l_{f,v,j}=k-1$ if $f(v,j)\in
    \{u_{0},\ldots ,u_{k-1}\} $ and $l_{f,v,j}=(k-1)^2$ 
    otherwise.  The value $l'_{g,v,j}$ is the same.
    Since only one of $f(v,j)$, $g(v,j)$ can be
    in  $\{u_{0},\ldots ,u_{k-1}\} $ we have $
    l_{\max }=O(k^3)$ and the lower bound follows.
  \end{proof}

  For the matrix model, there is a much simpler lower bound 
  which works even for undirected graphs.

  \begin{theorem}				\label{thm:2cycles}
    \textsc{Connectivity} needs $ \Omega (n^{3/2}) $ queries in
    the matrix model.
  \end{theorem}
  \begin{proof}
    We use Ambainis' method for the following
    special problem. You are given a symmetric
    matrix $ M\in \{0,1\}^{n\times n}$
    with the promise that it is the adjacency matrix of a graph with
    exactly one or two cycles, and have to find out which is the case.

    Let $X$ be the set of all adjacency matrices of a unique
    cycle, and $Y$ the set of all adjacency matrices with exactly
    two cycles each of length between $n/3$ and $2n/3$. We define
    the relation $R\subseteq X\times Y$ as $M\;R\;M'$ if there
    exist $ a,b,c,d\in [n] $ such that the only difference between
    $M$ and $M'$ is that $(a,b), (c,d)$ are edges in $M$ but not in $M'$
    and $(a,c), (b,d)$ are edges on $M'$ but not in $M$. 
    See figure~\ref{fig:cycles}.
    The definition of $Y$ implies that in $M$ the
    distance from $a$ to $c$ is between $ n/3 $ and $ 2n/3 $.

    \begin{figure}[ht]
      \centerline{\epsfig{file=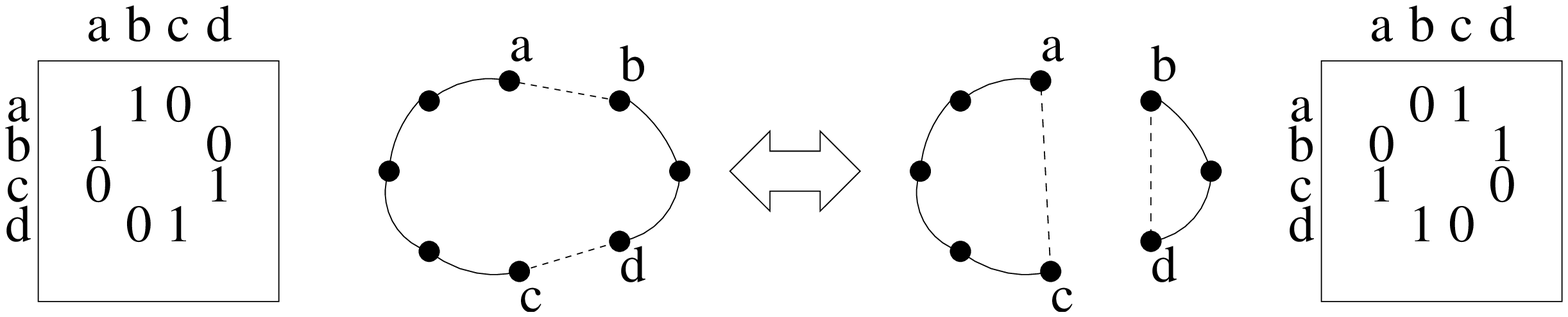,width=10cm}}
      \caption{Illustration of the relation}
      \label{fig:cycles}
    \end{figure}

    Then $m=O(n^{2})$ since there are $n-1$ choices
    for the first edge and $n/3$ choices for the second edge. 
    Also $m'=O(n^2)$ since from each cycle one edge must be picked,
    and cycle length is at least $n/3$.

    We have $l_{M,(i,j)}=4$ if $M_{i,j}=0$
    since in $M'$ we have the additional edge $(i,j)$ and
    the endpoints of the second edge must be neighbors of $i$ and $j$
    respectively.
    Moreover $l_{M,(i,j)}=O(n)$
    if $M_{i,j}=1$ since then $(i,j)$ is one of the edges to be removed
    and there remains $n/3$ choices for the second edge.

    The values $l'_{M',(i,j)}$ are similar,
    so in the product one factor will always be constant while
    the other is linear giving $l_{M,(i,j)}l'_{M',(i,j)}=O(n)$ and
    the theorem follows.
  \end{proof}

  \section{Conclusion}

  It remains to close the gap between lower and upper bound for
  regular directed graphs, for which we have
  only the bounds $\Omega(n)$ and $O(n^{3/2}\log n)$. 
  Our upper bounds relied
  only on Grover's algorithm.  One might hope for some cases that
  quantum random walks provide better upper bounds. 
  On the other side one might obtain 
  better lower bounds 
  with the polynomial method, which has recently be smartly used by
  Aaronson~\cite{Aaro02} and Shi~\cite{Shi02} to show lower bounds for
  properties of function graphs.

  \section*{Acknowledgments}

  We are grateful to Miklos Santha for helpful discussions, to Katalin
  Friedl for helpful comments on Algorithm~\ref{alg:learning},
  to Oded Regev for helpful corrections and to
  Ronald de Wolf who made us aware of the problem of error propagation
  in repeated use of Grover's search.

\end{document}